\documentclass[10pt,twoside,twocolumn,english,aps,manuscript,aps,manuscript,preprint,showpacs,superscriptaddress]{revtex4}
\usepackage[T1]{fontenc}
\usepackage[latin9]{inputenc}
\usepackage{babel}
\usepackage{amstext}
\usepackage{graphicx}
\usepackage[unicode=true,pdfusetitle,
 bookmarks=true,bookmarksnumbered=false,bookmarksopen=false,
 breaklinks=false,pdfborder={0 0 1},backref=false,colorlinks=false]
 {hyperref}
\usepackage{breakurl}

\makeatletter

\providecommand{\tabularnewline}{\\}

\@ifundefined{textcolor}{}
{%
 \definecolor{BLACK}{gray}{0}
 \definecolor{WHITE}{gray}{1}
 \definecolor{RED}{rgb}{1,0,0}
 \definecolor{GREEN}{rgb}{0,1,0}
 \definecolor{BLUE}{rgb}{0,0,1}
 \definecolor{CYAN}{cmyk}{1,0,0,0}
 \definecolor{MAGENTA}{cmyk}{0,1,0,0}
 \definecolor{YELLOW}{cmyk}{0,0,1,0}
}

\makeatother

\makeatother

\begin{document}

\title{Ba$_{2}$YIrO$_{6}$ : A cubic double perovskite material with Ir$^{5+}$
ions}

\author{T. Dey}

\email[Email: ]{tusdey@gmail.com}

\altaffiliation[Present address: ]{Experimental Physics VI, EKM, University of Augsburg, 86159 Augsburg, Germany}

\affiliation{Leibniz Institute for Solid State and Materials Research IFW, Institute
for Solid State Research, 01069 Dresden, Germany}

\author{A. Maljuk}

\affiliation{Leibniz Institute for Solid State and Materials Research IFW, Institute
for Solid State Research, 01069 Dresden, Germany}

\author{D. V. Efremov}

\affiliation{Leibniz Institute for Solid State and Materials Research IFW, Institute
for Solid State Research, 01069 Dresden, Germany}

\author{O. Kataeva}

\affiliation{Leibniz Institute for Solid State and Materials Research IFW, Institute
for Solid State Research, 01069 Dresden, Germany}

\affiliation{Kazan Federal University, Kremlevskaya Strasse 18, 420008, Kazan,
Russia}

\affiliation{A.E. Arbuzov Institute of Organic and Physical Chemistry, Russian
Academy of Sciences, Arbuzov Strasse 8, Kazan 420088, Russia}

\author{S. Gass}

\affiliation{Leibniz Institute for Solid State and Materials Research IFW, Institute
for Solid State Research, 01069 Dresden, Germany}

\author{C. G. F. Blum}

\affiliation{Leibniz Institute for Solid State and Materials Research IFW, Institute
for Solid State Research, 01069 Dresden, Germany}

\author{F. Steckel}

\affiliation{Leibniz Institute for Solid State and Materials Research IFW, Institute
for Solid State Research, 01069 Dresden, Germany}

\author{D. Gruner}

\affiliation{Leibniz Institute for Solid State and Materials Research IFW, Institute
for Solid State Research, 01069 Dresden, Germany}

\author{T. Ritschel}

\affiliation{Leibniz Institute for Solid State and Materials Research IFW, Institute
for Solid State Research, 01069 Dresden, Germany}

\author{A. U. B. Wolter}

\affiliation{Leibniz Institute for Solid State and Materials Research IFW, Institute
for Solid State Research, 01069 Dresden, Germany}

\author{J. Geck}

\affiliation{Leibniz Institute for Solid State and Materials Research IFW, Institute
for Solid State Research, 01069 Dresden, Germany}

\author{C. Hess}

\affiliation{Leibniz Institute for Solid State and Materials Research IFW, Institute
for Solid State Research, 01069 Dresden, Germany}

\affiliation{Center for Transport and Devices of Emergent Materials, TU Dresden,
01069 Dresden, Germany}

\author{K. Koepernik}

\affiliation{Leibniz Institute for Solid State and Materials Research IFW, Institute
for Solid State Research, 01069 Dresden, Germany}

\author{J. van den Brink}

\affiliation{Leibniz Institute for Solid State and Materials Research IFW, Institute
for Solid State Research, 01069 Dresden, Germany}

\affiliation{Institute for solid state physics, Technische Universität Dresden,
D-01062 Dresden, Germany}

\author{S. Wurmehl}

\affiliation{Leibniz Institute for Solid State and Materials Research IFW, Institute
for Solid State Research, 01069 Dresden, Germany}

\affiliation{Institute for solid state physics, Technische Universität Dresden,
D-01062 Dresden, Germany}

\author{B. Büchner}

\affiliation{Leibniz Institute for Solid State and Materials Research IFW, Institute
for Solid State Research, 01069 Dresden, Germany}

\affiliation{Center for Transport and Devices of Emergent Materials, TU Dresden,
01069 Dresden, Germany}

\affiliation{Institute for solid state physics, Technische Universität Dresden,
D-01062 Dresden, Germany}
\begin{abstract}
Materials with a $5d^{4}$ electronic configuration are generally
considered to have a nonmagnetic ground state ($J=0$). Interestingly,
Sr$_{2}$YIrO$_{6}$ (Ir$^{5+}$ having $5d^{4}$ electronic configuration)
was recently reported to exhibit long-range magnetic order at low
temperature and the distorted IrO$_{6}$ octahedra were discussed
to cause the magnetism in this material. Hence, a comparison of structurally
distorted Sr$_{2}$YIrO$_{6}$ with cubic Ba$_{2}$YIrO$_{6}$ may
shed light on the source of magnetism in such Ir$^{5+}$ materials
with $5d^{4}$ configuration. Besides, Ir$^{5+}$ materials having
$5d^{4}$ are also interesting in the context of recently predicted
excitonic types of magnetism. Here we report a single-crystal-based
analysis of the structural, magnetic, and thermodynamic properties
of Ba$_{2}$YIrO$_{6}$. We observe that in Ba$_{2}$YIrO$_{6}$ for
temperatures down to 0.4~K, long-range magnetic order is absent but
at the same time correlated magnetic moments are present. We show
that these moments are absent in fully relativistic \textit{ab initio}
band-structure calculations; hence, their origin is presently unclear.
\end{abstract}


\maketitle

\section{Introduction}

The iridates have become an interesting playground for material researchers
as they show novel ground states due to competing interactions between
the crystal field (CF), the Coulomb interaction ($U$) and the spin-orbit
coupling (SOC) \cite{Cao book}. To name a few examples, (Sr/Ba)$_{2}$IrO$_{4}$
\cite{Kim-PRL-101-2008,Kim-Science-323-2009,Katukuri-PRB-85-2012,Okabe-PRB-83-2011,Boseggia-PRL-110-2013},
Sr$_{3}$Ir$_{2}$O$_{7}$ \cite{Fujiyama-PRB-86-2012-Sr3Ir2O7,Cao-PRB-66-2002-Sr3Ir2O7,Carter-PRB-87-2013},
(Na/Li)$_{2}$IrO$_{3}$ \cite{Singh-PRL-108-2012,Singh-PRB-2010-Na2IrO3,Manni-PRB-2014,Chaloupka-PRL-2010,Price-PRL-109-2012,Gretarsson-PRL-110-2013}
have been studied intensively in recent times, both experimentally
and theoretically.

\begin{figure*}
\centering{}\includegraphics[scale=0.13]{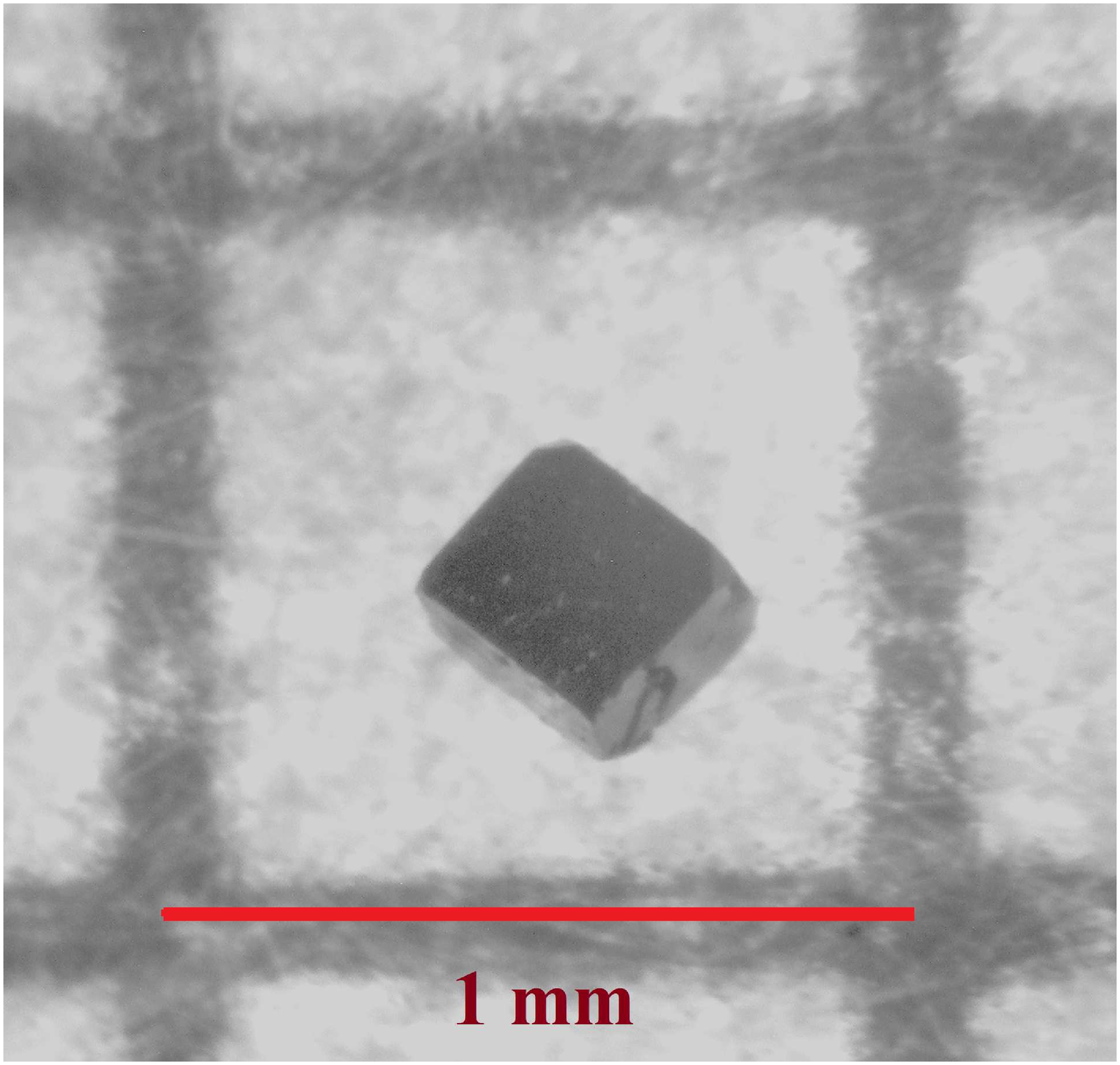} \includegraphics[scale=0.185]{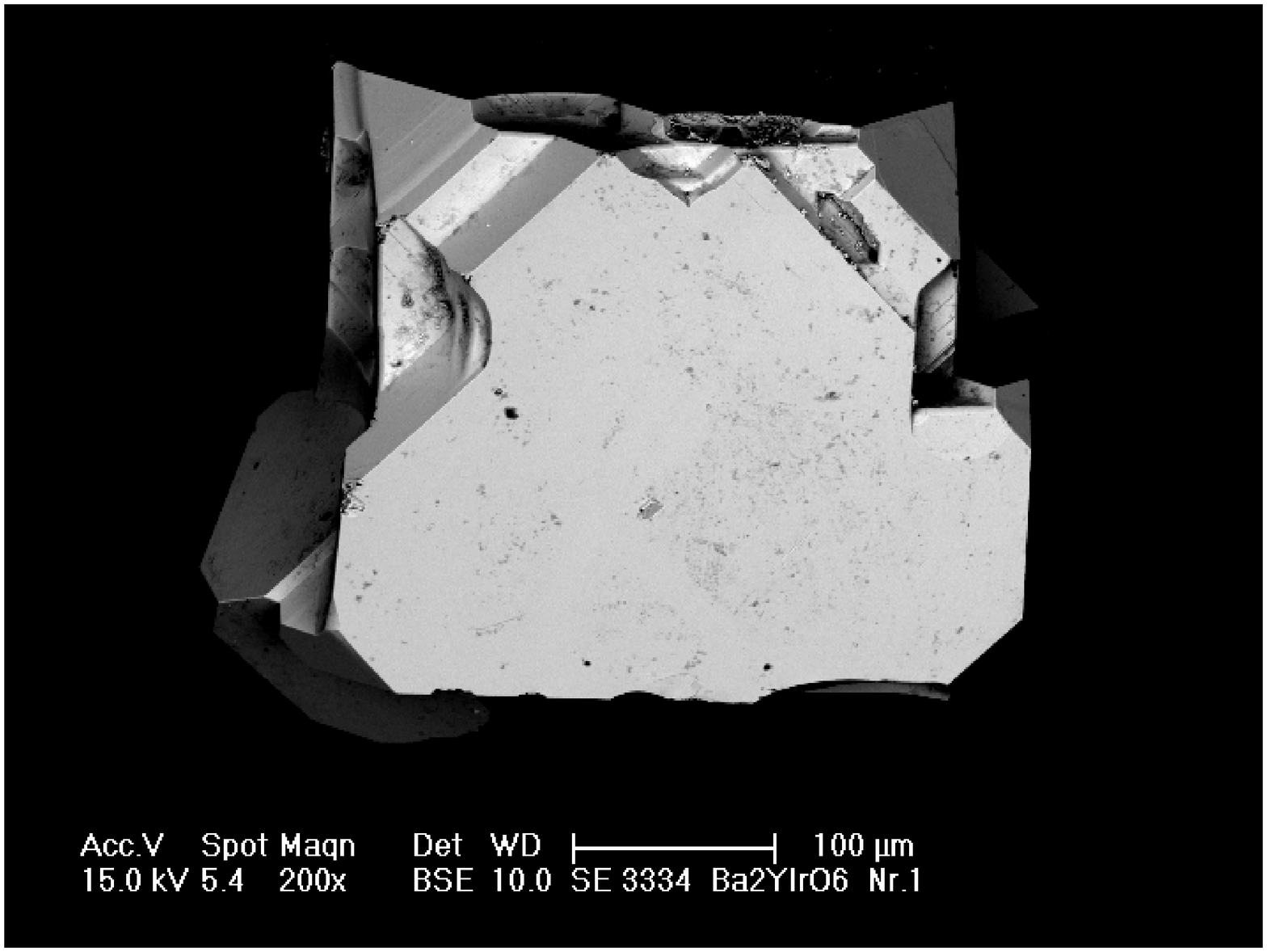}
\includegraphics[scale=0.15]{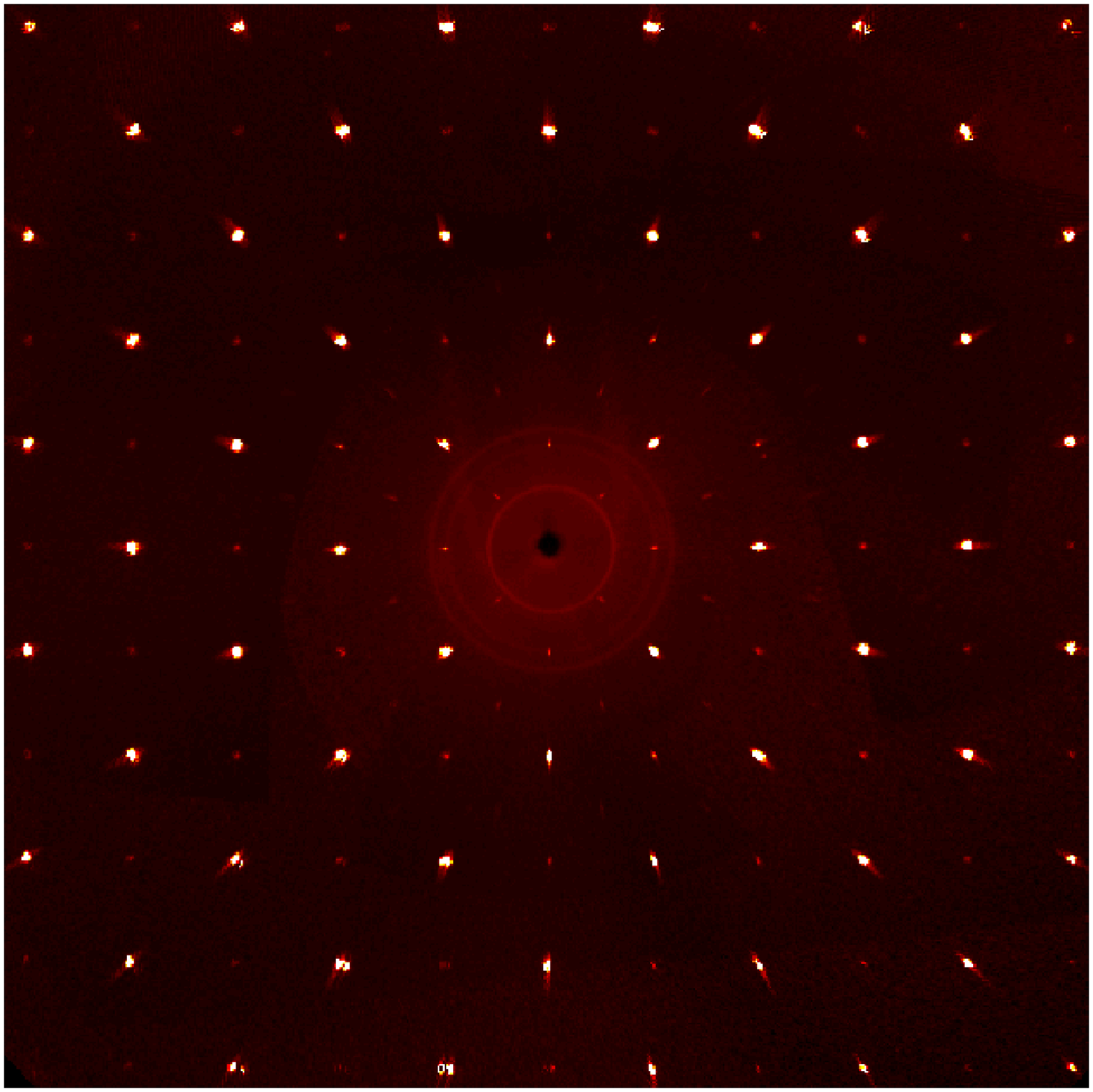}\protect\protect\caption{\label{fig:SEMpic} (Left) Optical image of a Ba$_{2}$YIrO$_{6}$
single crystal placed on a millimeter graph paper. (Middle) SEM image
of one Ba$_{2}$YIrO$_{6}$ single crystal in backscattering electron
(BSE) mode. (Right) A representative single-crystal XRD pattern in
the (h0l) plane at $198(2)$K.}
\end{figure*}

\begin{figure}
\centering{}\centering{}\includegraphics[scale=0.3]{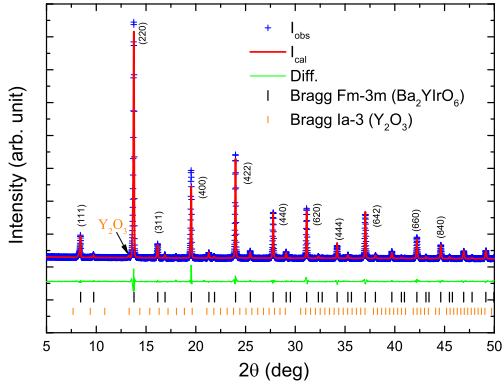}\protect\protect\caption{\label{fig:PowderXRD} Powder XRD pattern (blue cross) and its double-phase
refinement (red solid line); space groups $Fm\overline{3}m$ for Ba$_{2}$YIrO$_{6}$
and $Ia\overline{3}$ for Y$_{2}$O$_{3}$ are shown. The Bragg positions
corresponding to space groups $Fm\overline{3}m$ and $Ia\overline{3}$
are shown as black and orange vertical lines, respectively. The green
line shows the difference between the observed and the calculated
patterns. The main reflection corresponding to Y$_{2}$O$_{3}$ is
marked with an arrow.}
\end{figure}

In all the above mentioned compounds, the iridium ion is magnetic
with a formal oxidation state $+4$ ($5d$$^{5}$). In contrast, materials
with a $4d^{4}$ or $5d^{4}$ electronic configuration (such as Re$^{3+}$,
Ru$^{4+}$, Os$^{4+}$, Ir$^{5+}$) are believed to be in a Van Vleck-type
non-magnetic band insulating ground state with a completely filled
$J=3/2$ manifold having a total angular momentum $J=0$ \cite{Chen-Balents-PRB-2011-d4}.
Such a nonmagnetic ground state is realized in NaIrO$_{3}$ \cite{Bremholm-JSSC-184-2011,Du-EPL-2013-NaIrO3}.

However, recently two independent theoretical studies proposed that
the interplay between $U$ and SOC in some of these materials may
lead to novel magnetism governed by gapped singlet-triplet excitations
\cite{Khaliullin-PRL-2013-d4magnetism,Meetei-arxiv-2013-d4magnetism},
although the ground states obtained in these two studies are different.
Meetei \textit{et al.} \cite{Meetei-arxiv-2013-d4magnetism} proposed
a magnetic phase diagram for the $d^{4}$ Mott insulators which consists
of a nonmagnetic ($J=0$) and two different ferromagnetic phases ($J=2$
and $J=1$). The authors have identified the double perovskite materials
as good candidates to observe such novel magnetic states.

Recently, in the double perovskite material Sr$_{2}$YIrO$_{6}$ with
Ir$^{5+}$ ions, a transition to an antiferromagnetic long-range order
was observed below $1.3$~K \cite{Cao-PRL-2014-Sr2YIrO6}. Cao \textit{et
al.} \cite{Cao-PRL-2014-Sr2YIrO6} assigned the structural distortion
of Sr$_{2}$YIrO$_{6}$ originating in the monoclinic structure (space
group $P2_{1}/n$) with highly distorted IrO$_{6}$ octahedra as the
driving force for the long-range magnetic order in this compound.
Hence, the motivation to study the Ba analog Ba$_{2}$YIrO$_{6}$
is twofold: (i) the investigation of other Ir$^{5+}$ materials in
general to verify or discard the theoretically predicted excitonic
magnetism in the Van Vleck-type $d^{4}$ Mott insulators \cite{Khaliullin-PRL-2013-d4magnetism,Meetei-arxiv-2013-d4magnetism},
and (ii) the investigation of cubic analogs with Ir$^{5+}$ to shed
light on the impact of noncubic symmetry on the magnetism in such
systems.

The crystal structure of Ba$_{2}$YIrO$_{6}$ has been described using
a cubic symmetry (space group $Fm\overline{3}m$) \cite{Choy-JACS-1995,Thumm-JSSC-1980,Fu-JAlloyComp-394-2005}
as well as a monoclinic symmetry (space group \textit{$P2_{1}/n$})
with $\beta=90.039\text{º}$ (Ref.~ \onlinecite{Wakeshima-JAlloyComp-287-1999}).
The monoclinic description with $\beta\sim90\text{º}$ closely matches
with the cubic description. In any case, since all previous studies
are based on polycrystalline samples, it is important to revisit the
crystal structure of Ba$_{2}$YIrO$_{6}$ using single crystals and
in particular shedding light on the magnetic properties of this material.

In this work, we report the growth of Ba$_{2}$YIrO$_{6}$ single
crystals and the details of their structural, magnetic, thermal, and
electrical transport properties. Ba$_{2}$YIrO$_{6}$ clearly crystallizes
in a cubic double perovskite-type (space group $Fm\overline{3}m$)
structure, as demonstrated by our single-crystal and powder (on crushed
single crystals) x-ray diffraction study, and, hence, is a suitable
material to address the open issues in Ir$^{5+}$ materials as outlined
above. The crystals are semiconducting and show a paramagnetic behavior
in the temperature range $0.43-300$ K. Our susceptibility and heat
capacity measurements show no sign of long-range magnetic ordering
down to at least 0.4 K. The effective magnetic moment ($\mu_{eff}=0.44$
$\mu_{B}$/Ir) obtained from the Curie-Weiss fit of our susceptibility
data is unexpected for the anticipated $J=0$ material. This could
be arising from chemical disorder in the crystals. We performed \textit{ab
initio} calculations in the LDA+U scheme to gain insight into the
ground state. The results show that the initially metallic band becomes
insulating due to the interplay of the spin-orbit and the Coulomb
interaction.

\begin{figure}
\centering{}\centering{}\includegraphics[scale=0.9]{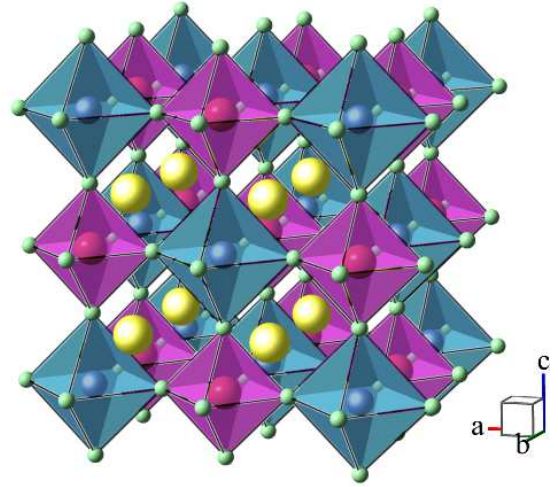}\protect\protect\caption{\label{fig:Structure} Crystal structure of Ba$_{2}$YIrO$_{6}$.
The yellow and the green atoms represent Ba and O, respectively. The
pink and light blue octahedra correspond to the IrO$_{6}$ and YO$_{6}$
octahedra, respectively.}
\end{figure}

\begin{table}
\protect\protect\caption{\label{tab:SingleCrystalXRD} Crystal data for Ba$_{2}$YIrO$_{6}$
from single-crystal diffractometry.}

\centering{}%
\begin{tabular}{|c|c|}
\hline 
Temperature (K)  & \textbf{$198(2)$}\tabularnewline
\hline 
\hline 
Crystal (for XRD) size (mm$^{3}$)  & $0.15\times0.16\times0.19$\tabularnewline
\hline 
Space group  & $Fm\overline{3}m$ (No. $225$)\tabularnewline
\hline 
$a$ ($\textrm{\AA}$)  & $8.3387(8)$\tabularnewline
\hline 
V ($\textrm{\AA}^{3}$)  & $579.8(2)$\tabularnewline
\hline 
$Z$ & $4$\tabularnewline
\hline 
$\rho_{calc}$(g cm$^{-3}$)  & $7.467$\tabularnewline
\hline 
$\mu$(mm$^{-1}$)  & $46.149$\tabularnewline
\hline 
Multiscan absorption correction  & $0.041\le T\le0.055$\tabularnewline
\hline 
$\theta$ range ($^{\circ}$)  & $0.95-45.1$\tabularnewline
\hline 
Collected reflections  & $12012$\tabularnewline
\hline 
Independent reflections  & $162(R_{int}=0.0486)$\tabularnewline
\hline 
Observed reflections  & $162[I\ge2\sigma(I)]$\tabularnewline
\hline 
Refined parameters  & $8$\tabularnewline
\hline 
$R$  & $0.0096$\tabularnewline
\hline 
$wR^{2}$  & $0.0270[I>2\sigma(I)]$\tabularnewline
\hline 
Max residual electron density ($\textrm{\ensuremath{e}\AA}^{-3}$)  & $1.459$\tabularnewline
\hline 
Min residual electron density ($\textrm{\ensuremath{e}\AA}^{-3}$)  & $-0.559$\tabularnewline
\hline 
Goodness of fit  & $1.350$\tabularnewline
\hline 
\end{tabular}
\end{table}

\begin{table}
\centering{}\protect\protect\caption{\label{tab:AtomicPositions} Atomic positions and thermal parameters
for a Ba$_{2}$YIrO$_{6}$ single crystal at $198(2)$ K.}

\begin{tabular}{|c|c|c|c|c|c|c|}
\hline 
Atoms  & Site  & x  & y  & z  & Site occ.  & U$_{iso}$\tabularnewline
\hline 
Ba(1)  & 8c  & 0.25  & 0.25  & 0.25  & 1  & 1.0675\tabularnewline
\hline 
Y(1)  & 4a  & 0  & 0  & 0  & 1  & 0.75887\tabularnewline
\hline 
Ir(1)  & 4b  & 0.5  & 0.5  & 0.5  & 1  & 0.81458\tabularnewline
\hline 
O(1)  & 24e  & 0.25944(2)  & 0  & 0  & 1  & 0.51547\tabularnewline
\hline 
\end{tabular}
\end{table}

\section{Experimental Details}

Single crystals of Ba$_{2}$YIrO$_{6}$ were grown using high-purity
starting materials BaCO$_{3}$ (Alpha Aesar $99.997\%$), IrO$_{2}$
(Alpha Aesar $99.99\%$), and Y$_{2}$O$_{3}$ (Alpha Aesar $99.999\%$).
Ultra-dry BaCl$_{2}$ (Alpha Aesar $99.5\%$) was used as flux. The
stoichiometric mixture of the starting materials and the flux were
put inside a platinum crucible covered with a platinum lid to reduce
flux evaporation. The mixture was heated to $1250-1300\text{º}$C,
held at this temperature for $24$ h and then slowly cooled to $950\text{º}$C.
After that it was cooled to room temperature very fast by switching
off the furnace. Cubic-size single crystals (typical dimensions $0.3\times0.3\times0.3$
mm$^{3}$, also compare Fig. \ref{fig:SEMpic}) were precipitated
at the bottom of the platinum crucible. After dissolving the flux
in water, the single crystals were collected and used for further
characterization.

Single-crystal x-ray diffraction (XRD) data of Ba$_{2}$YIrO$_{6}$
were collected on a Bruker AXS Kappa APEX diffractometer with graphite-monochromated
Mo-$K\alpha$ radiation ($\lambda=0.71073\textrm{\AA}$) using the
complete sphere mode at $198(2)$ K. The following programs were used
to collect and analyze the data: the data collection was done with
APEX2 \cite{Bruker-APEX2}, the data reduction with `SAINT' \cite{Bruker-Areadetector},
the numerical absorption correction was applied using SADABS \cite{Sheldrick-Gottingen-1996},
the structure solution was obtained with SHELXS-97 \cite{Sheldrick-ActCryst-1993},
and the structure refinement was performed by full-matrix least-squares
against $F^{2}$ using SHELXL-97 \cite{Sheldrick-ActCryst-1993}.
CSD 427064 contains the supplementary crystallographic data for this
work. Room temperature powder XRD patterns were obtained using a Stoe
Stadi-P diffractometer with Mo-$K\alpha1$ radiation equipped with
a curved Ge($111$) primary monochromator and a Dectris Mythen 1 K
detector. Our data were analyzed with the Rietveld method using the
FULLPROF program \cite{Rodriguez-Carvajal-PhysicaB-1993}. The homogeneity
and the chemical composition of the crystals were analyzed using energy-dispersive
x-ray (EDX) analysis with a scanning electron microscope (SEM Philips
XL 30).

Magnetization measurements were performed as a function of temperature
($T$) and magnetic field ($H$) on randomly oriented single crystals
of a total mass $\sim$ 57 mg in a Quantum Design MPMS in the temperature
range $0.43-300$ K using the \textit{iHelium3} option below 1.8 K.
Heat capacity measurements ($C_{P}$) were performed on $10-12$ crystals
(mass $\sim$ 2 mg) in the temperature range $0.4-10$ K using the
$^{3}$He option of a Quantum Design PPMS. The resistivity was measured
as a function of temperature in the range $4.2-300$~K in a homemade
device. The contacts on the sample were made using graphite epoxy
in the four-point contact geometry.

\section{Results and discussions}

\subsection{Microstructure analysis}

Figure \ref{fig:SEMpic} exemplarily shows an as-grown Ba$_{2}$YIrO$_{6}$
single crystal (left panel). The middle panel of Fig. \ref{fig:SEMpic}
shows the SEM image of a single crystal in backscattered electron
(BSE) mode. Our SEM analysis suggests that the composition is homogeneous
over the crystals. However, we cannot exclude the possibility of a
certain amount of off-stoichiometry, e.g., Ba$_{2}$Y$_{1+d}$Ir$_{1-d}$O$_{6-d}$.

\begin{figure}
\centering{}\centering{}\includegraphics[scale=0.5]{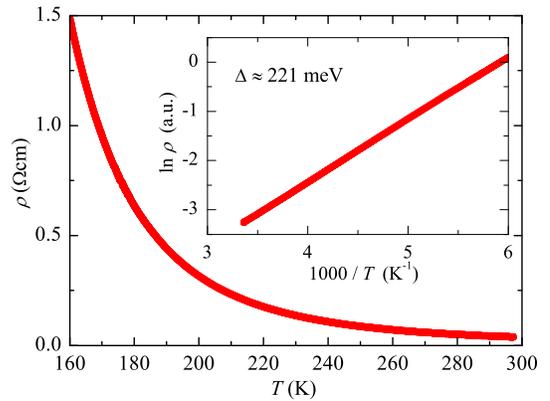}\protect\protect\caption{\label{fig:resistivity} Resistivity ($\rho$) as a function of temperature
for a single-crystal piece of Ba$_{2}$YIrO$_{6}$. Inset: Arrhenius
plot of the resistivity.}
\end{figure}

\begin{figure}
\centering{}\centering{}\includegraphics[scale=0.3]{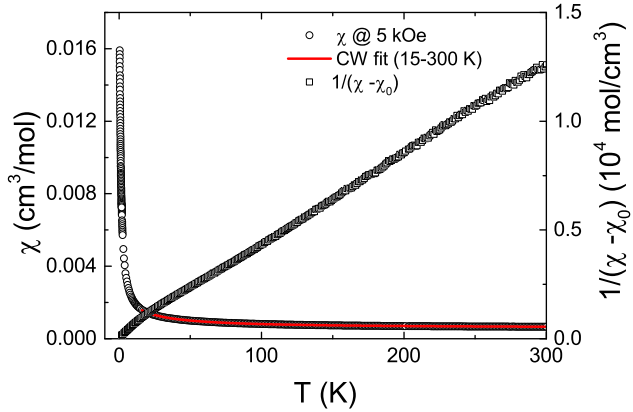}\protect\protect\caption{\label{fig:MvsT} (Left axis) Magnetic susceptibility in an external
magnetic field $H=0.5$~kOe of as-grown crystals as a function of
temperature along with its fit according to Curie-Weiss (CW) law in
the range $15-300$~K. (Right axis) The inverse susceptibility (after
subtracting $\chi_{0}$ obtained from the CW fit).}
\end{figure}

\subsection{XRD and crystal structure}

Single-crystal XRD measurements (shown in the right panel of Fig.
\ref{fig:SEMpic}) performed on several pieces from different preparation
batches showed a high quality of the crystals, proven by the good
internal consistency of the data collected using the full-sphere mode
and an extremely low $R$ factor (less than $1\%$). These measurements
confirm that Ba$_{2}$YIrO$_{6}$ crystallizes in a cubic double perovskite
structure with space group $Fm\overline{3}m$ (No. $225$), similar
to results as reported in Refs. \cite{Choy-JACS-1995,Fu-JAlloyComp-394-2005,Thumm-JSSC-1980}
based on XRD measurements on polycrystalline samples. The results
of the structural refinement of the single-crystal XRD measurements
at $198(2)$ K are summarized in Table \ref{tab:SingleCrystalXRD}.
The atomic positions and thermal parameters are listed in Table \ref{tab:AtomicPositions}.

Furthermore, we have measured powder XRD on crushed single crystals.
The resulting XRD pattern is shown in Fig. \ref{fig:PowderXRD}. Traces
of unreacted Y$_{2}$O$_{3}$ ($\sim2\%$) are found in the XRD pattern
with the main peak corresponding to Y$_{2}$O$_{3}$ marked by an
arrow along with small amounts of Pt from the crucible that are contaminating
the surfaces of the crystal. All the major peaks in the powder XRD
pattern are indexed with space group $Fm\overline{3}m$ as shown in
Fig. \ref{fig:PowderXRD}. The double-phase Rietveld refinement of
the powder XRD pattern (shown in Fig. \ref{fig:PowderXRD}) using
space group $Fm\overline{3}m$ for Ba$_{2}$YIrO$_{6}$ and $Ia\overline{3}$
for Y$_{2}$O$_{3}$ results in refinement parameters $R_{p}=2.54$
and $R_{wp}=3.88$. The lattice constants obtained from single-crystal
XRD and powder XRD on crushed crystals are consistent with earlier
reports \cite{Wakeshima-JAlloyComp-287-1999,Thumm-JSSC-1980,Choy-JACS-1995,Fu-JAlloyComp-394-2005}.
We find no signature of any structural transition of the crystals
from room temperature (powder XRD) down to $198(2)$ K (single-crystal
XRD). Also, no anomaly is seen in our magnetic susceptibility and
heat capacity measurement (discussed later), which suggests the absence
of any structural transition down to 0.4~K.

The crystal structure of Ba$_{2}$YIrO$_{6}$ based on our refinement
results is shown in Fig. \ref{fig:Structure}. An alternating arrangement
of IrO$_{6}$ (pink) and YO$_{6}$ (light blue) octahedra, with the
Ba atoms (yellow) placed in between, form the crystal structure. Please
note that the Ir$^{5+}$ ions in this structure build up a face-centered
cubic (fcc) network. Attempts to allow site disorder in the refinement
were not leading to better fit results; hence, we concluded that site
disorder does not play a role here. This is consistent with the ionic
size of Y$^{3+}$ and Ir$^{5+}$ ions being very different, rendering
site disorder rather unlikely. The same arguments hold for off-stoichiometry.
The distortion of IrO$_{6}$ octahedra as present in the monoclinic
Sr$_{2}$YIrO$_{6}$ is absent in the cubic Ba$_{2}$YIrO$_{6}$.

\begin{figure}
\centering{}\centering{}\includegraphics[scale=0.3]{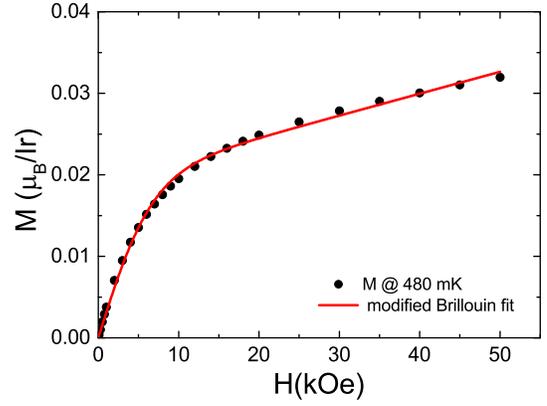}\protect\protect\caption{\label{fig:MvsH} The isothermal magnetization curve for as-grown
crystals at $T=480$~mK is shown together with a fit according to
a modified Brillouin function; for details see text.}
\end{figure}

\begin{figure}
\centering{}\centering{}\includegraphics[scale=0.28]{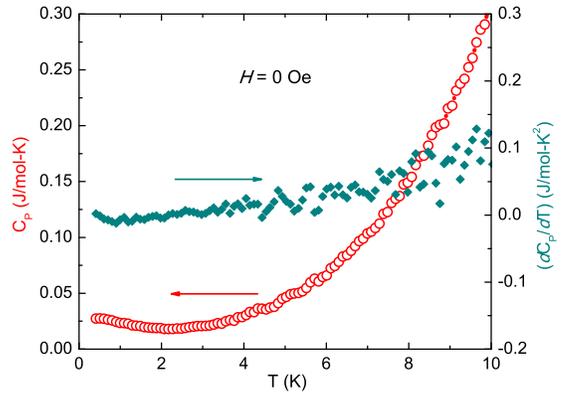}\protect\protect\caption{\label{fig:HC} (Left axis) The specific heat ($C_{P}$) of several
Ba$_{2}$YIrO$_{6}$ single crystals in zero field as a function of
temperature ($T$). (Right axis) The derivative of $C_{P}$.}
\end{figure}

\begin{figure*}
\includegraphics[scale=0.28]{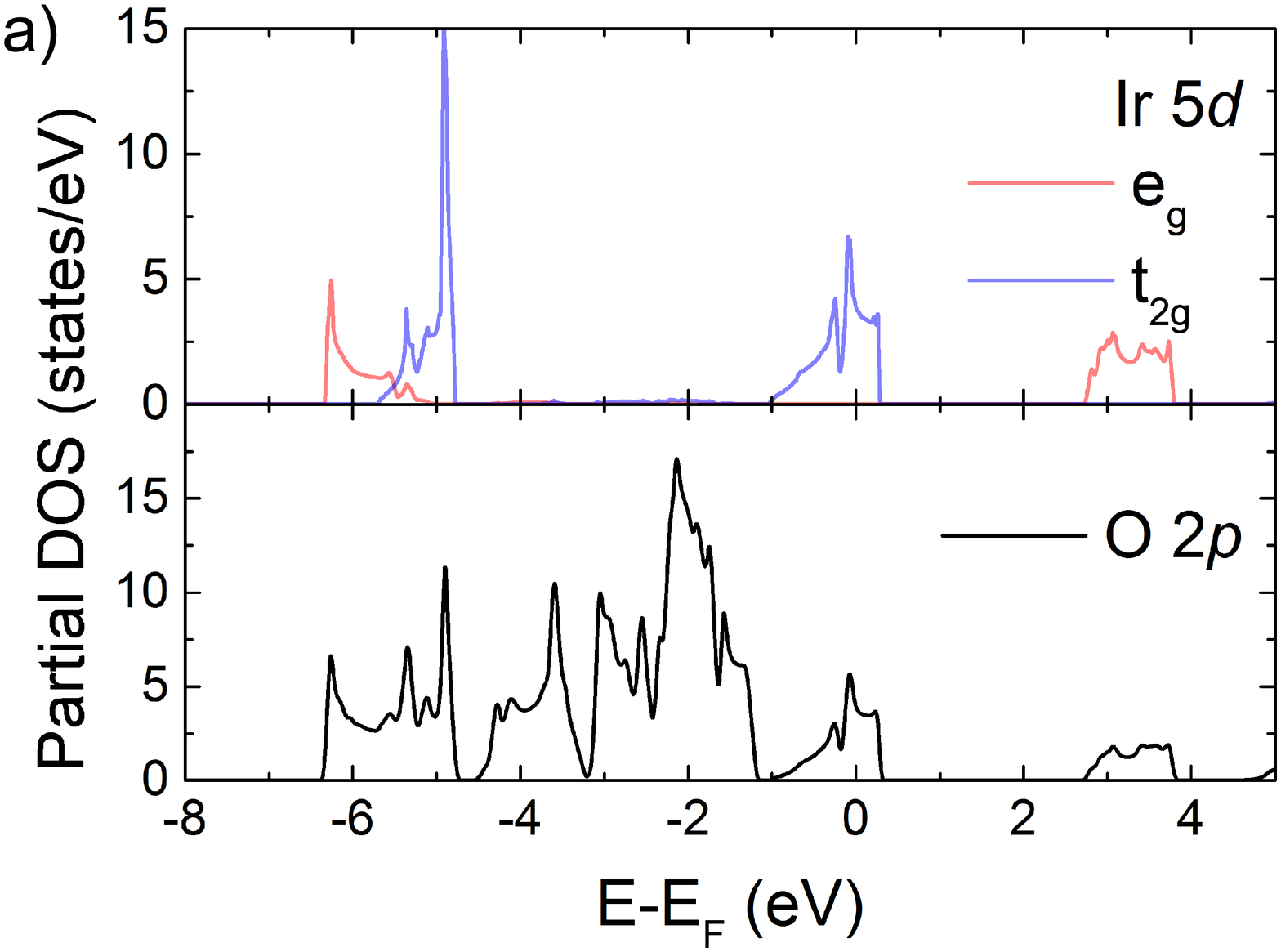} \includegraphics[scale=0.28]{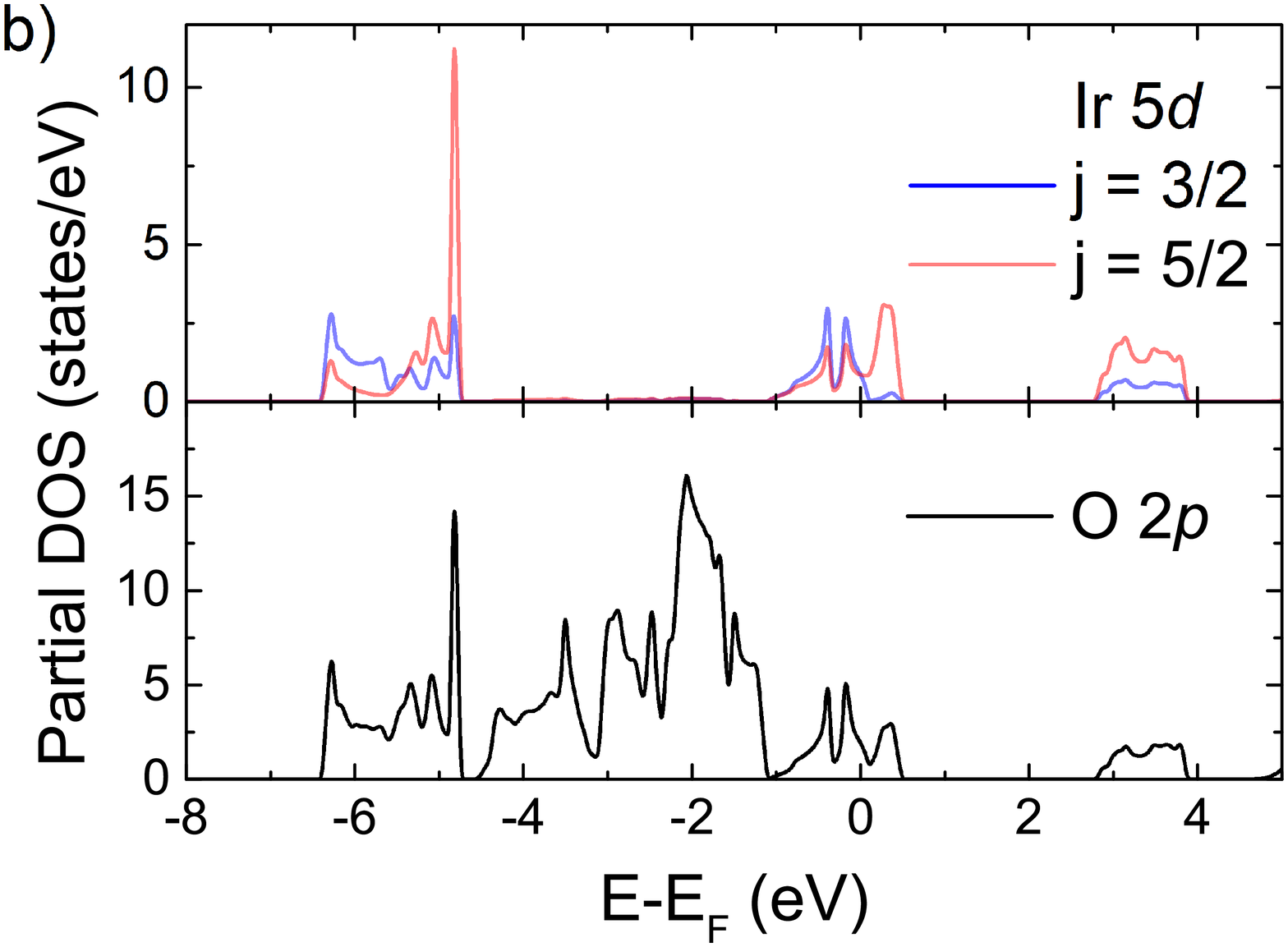}
\includegraphics[scale=0.28]{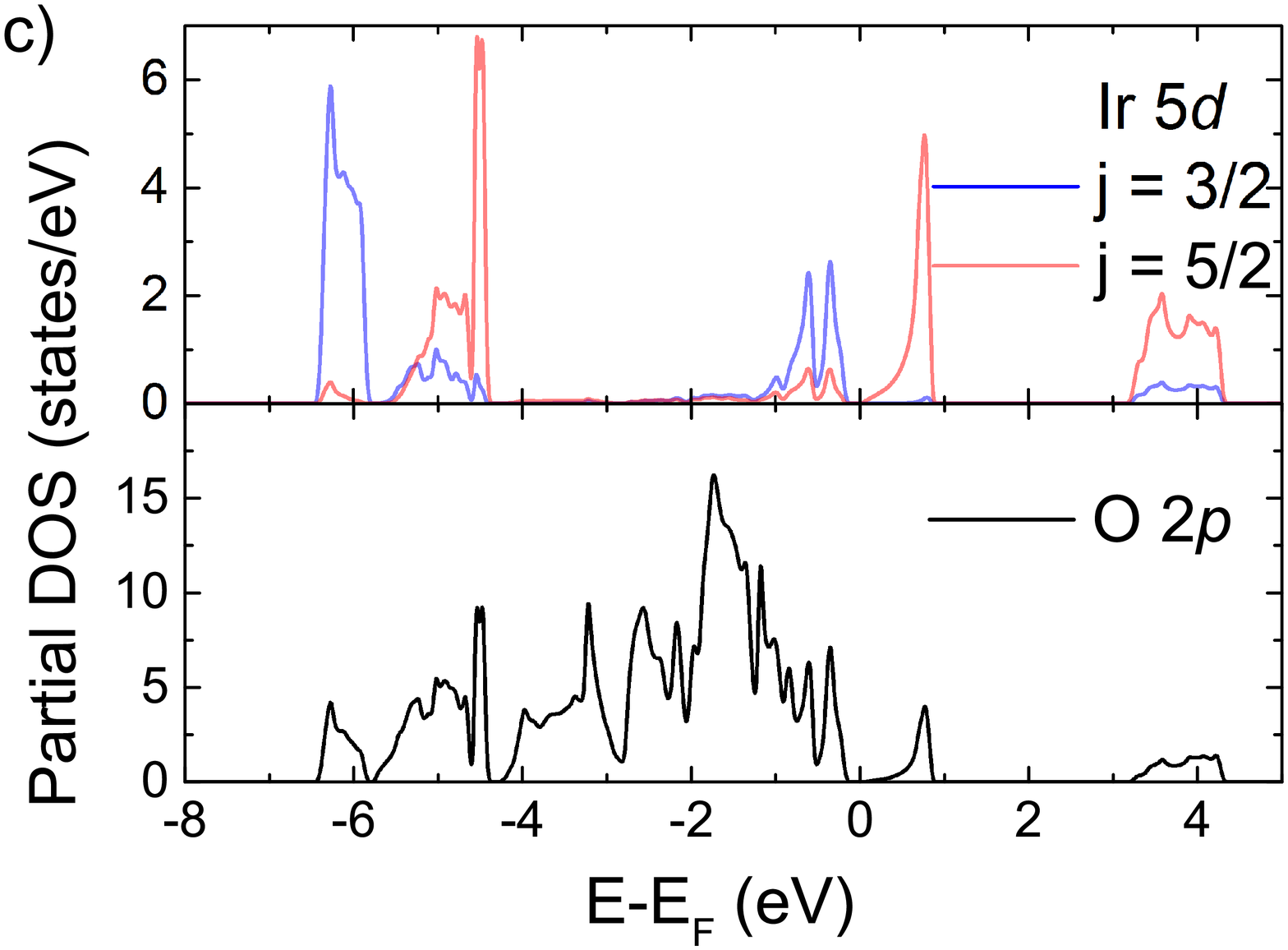} \includegraphics[scale=0.3]{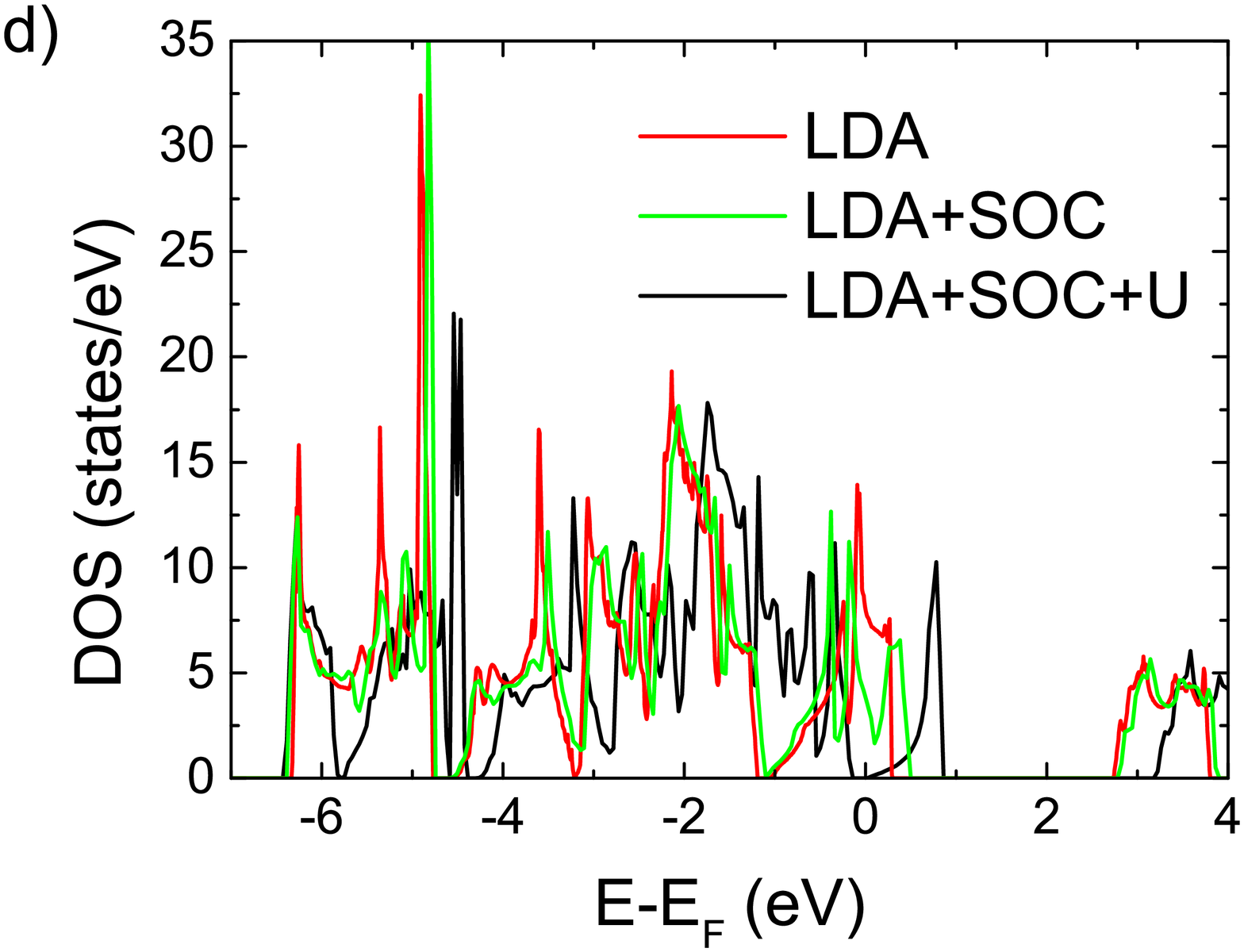}
\caption{\label{fig:pDOS} Partial density of states (pDOS) for Ir and O atoms:
(a) LDA scheme, (b) LDA+SOC scheme, (c) LDA+SOC+U scheme, (d) total
density of states, calculated in the frame of LDA, LDA+SOC, LDA+SOC+U.}
\end{figure*}

\subsection{Resistivity}

Figure \ref{fig:resistivity} shows the semiconductor-type resistivity
of a Ba$_{2}$YIrO$_{6}$ single crystal as a function of temperature.
The measurement was done at a constant current of $I=5~\mu$A. The
resistivity at room temperature is $\rho$($300$K)=$40$ m$\Omega$cm
and increases exponentially with decreasing temperature. At temperatures
lower than $\sim170$~K, the resistivity increases to very high values,
hindering a correct measurement with our device. The inset of Fig.
\ref{fig:resistivity} shows the Arrhenius plot of the resistivity
data. From room temperature down to $160$ K, ln($\rho$) is inversely
proportional to the temperature, i.e., $\rho\propto exp(\triangle/2k_{B}T)$.
Our analysis yields an energy gap of $\triangle\approx221$ meV. It
should be noted that the geometrical error of the contacts is quite
high due to the small sample size. This may influence the accuracy
of the absolute value of the resistivity; however, the overall temperature
dependence and therefore the energy gap is unaffected.

\subsection{Magnetization}

Since the magnetization of the material is expected to be small, we
have taken special care to increase the sample mass and to subtract
the background signal of the sample holder. Figure \ref{fig:MvsT}
shows the zero-field-cooled susceptibility data as a function of temperature
in an external magnetic field of 5~kOe for our as-grown single crystals.
No signature of any long-range magnetic order is found in the measured
temperature range $0.43$~K$\le T\le300$~K. Moreover, no splitting
between zero-field-cooled (ZFC) and field-cooled (FC) susceptibilities
are observed between 1.8 and 300~K, even when measured in small applied
fields of $50$~Oe \cite{note1}. The susceptibility data is fitted
with a Curie-Weiss law $\chi(T)=\chi_{0}+C/(T-\theta)$ in the temperature
range $15-300$~K (shown in Fig. \ref{fig:MvsT}). This fitting gives
a temperature independent susceptibility contribution $\chi_{0}=5.83\times10^{-4}$\ cm$^{3}$/mol,
a Curie constant $C=0.0247$\ cm$^{3}$K/mol (effective magnetic
moment $\mu_{eff}=0.44$\ $\mu_{\mathrm{B}}$/Ir), and a Weiss temperature
$\theta\sim-8.9$~K. The inverse susceptibility (after subtracting
$\chi_{0}$) is plotted on the right axis of the figure. Below $\sim15$~K,
deviations from the CW fitting occur, which probably stem from even
larger antiferromagnetic spin correlations in the low-temperature
regime, or from a small temperature-dependent contribution to the
Van Vleck susceptibility, which has not been taken into account in
the fit but which has been observed, e.g., for Eu$^{3+}$ ~\cite{Takikawa}.

Since materials with $5d^{4}$ electronic configuration are expected
to be Van Vleck-type nonmagnetic, it is interesting to extract the
Van Vleck part in the susceptibility for this material. For insulating
materials one can consider $\chi_{0}=\chi_{core}+\chi_{vv}$, where
$\chi_{core}$ is the core diamagnetic susceptibility and $\chi_{vv}$
is Van Vleck paramagnetic susceptibility. In case of Ba$_{2}$YIrO$_{6}$,
$\chi_{core}=-1.68\times10^{-4}$ cm$^{3}$K/mol (obtained by adding
the core diamagnetic susceptibility for individual ions \cite{Selwood-magnetochemistry}),
which results in $\chi_{vv}=7.51\times10^{-4}$ cm$^{3}$K/mol. This
value is of the same order of magnitude as for other Ir$^{5+}$ ($5d^{4}$)
materials (see Table \ref{tab:VanVleckSus}).

\begin{figure*}
\includegraphics[scale=0.3]{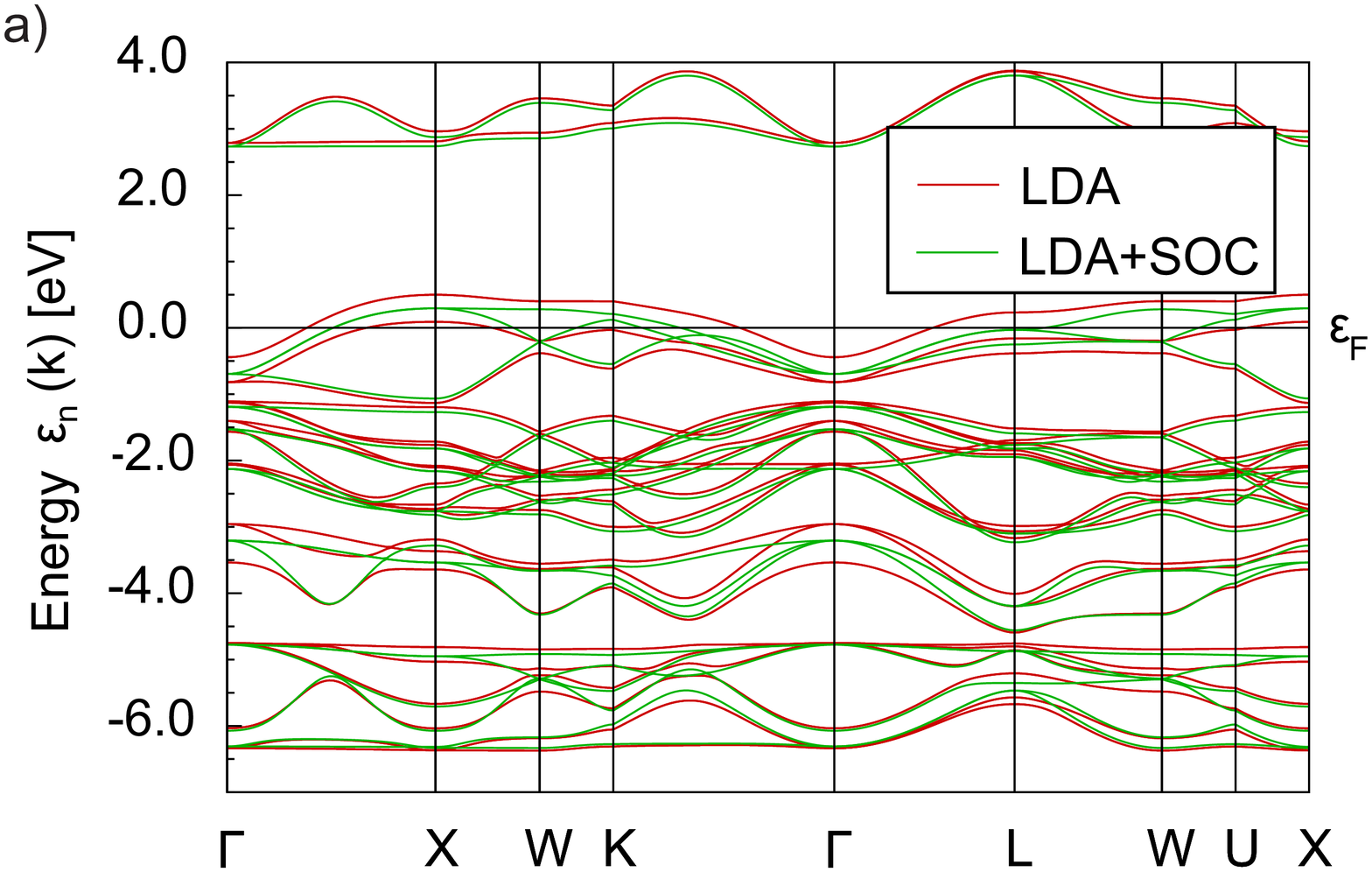} \includegraphics[scale=0.3]{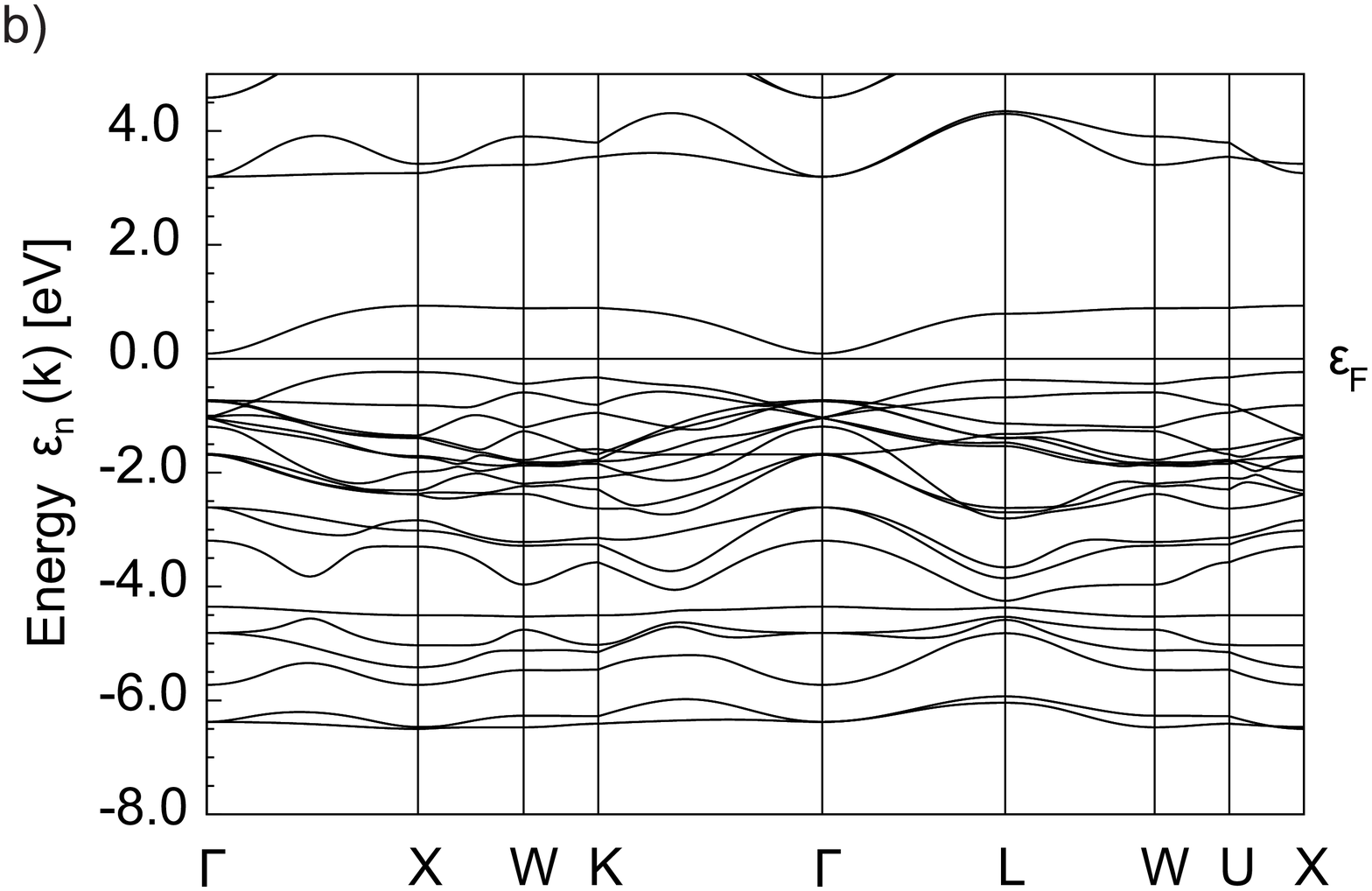}
\protect\caption{\label{fig:bands} Band structure: (a) LDA and LDA+SOC (b) LDA+SOC+U.}
\end{figure*}

The effective magnetic moment ($\mu_{eff}=0.44$\ $\mu_{\mathrm{B}}$/Ir)
is unusual for an expected $J=0$ material. The simplest explanation
for the magnetic response could be the presence of a few percent of
Ir$^{4+}$ ions (which are known to be magnetic) caused by oxygen
deficiency. To verify or discard the presence of oxygen vacancies,
we have annealed the as-grown crystals under $700$ bar oxygen pressure
at $500\text{º}$C for $2$ days. Magnetization measurements on these
oxygen-annealed crystals yield identical results compared with those
of the as-grown crystals. This probably suggests that there are no
oxygen vacancies in the as-grown crystals.

Another possibility for the presence of Ir$^{4+}$ ions is a partial
substitution of Y$^{3+}$ by Ir$^{4+}$ ions, as it was seen in the
cubic double perovskite material Ba$_{2}$(Y$_{0.67}$Ir$_{0.33}$)IrO$_{6}$
with the formal oxidation state of Ir $+4.5$ \cite{Dey-PRB-2013-Ba3YIr2O9}.
The small Curie constant of Ba$_{2}$YIrO$_{6}$ and Weiss temperature
could result from the presence of a few percent of Ir$^{4+}$ ions
($\sim5\%$ of $J=1/2$ spins). A third scenario accounts for Ir$^{6+}$
ions created due to the intermixing and/or off-stoichiometry. For
double perovskite materials with Ir$^{6+}$ ions, $\mu_{eff}$ is
reported to be more than 3\ $\mu_{\mathrm{B}}$/Ir ~\cite{Kayser-InorgChem-2013-Sr2ZnIrO6}.
Hence, the presence of a small amount of Ir$^{6+}$ ions can result
in the observed $\mu_{eff}=0.44$\ $\mu_{\mathrm{B}}$/Ir.

To understand the origin of the magnetic moments, we further analyzed
the isothermal magnetization data \textit{M}(\textit{H}) at 480~mK
as shown in Fig. \ref{fig:MvsH}. We were able to fit our data with
a modified Brillouin function $M(H)=\chi_{0}H+fN_{A}g\mu_{\mathrm{B}}JB_{J}(g\mu_{\mathrm{B}}JH/k_{B}T)$,
where $f$, $N_{A}$, $g$, and $k_{B}$ represent a scaling factor
to account for a finite number of paramagnetic impurities, the Avogadro
constant, the Landé $g$ factor, and the Boltzmann constant, respectively
(see Fig. \ref{fig:MvsH}). The first linear term $\chi_{0}H$ describes
the core plus the Van Vleck contributions (see above), while the second
term, the Brillouin function $B_{J}$, represents the behavior of
paramagnetic spins as a function of a magnetic field at a particular
temperature.

Our analysis suggests the presence of $\sim$2 \% of $J=1/2$ spins
in the material and a $g$ factor of 2.14. This result matches quite
well with the estimate from the CW fitting assuming $g$ = 2.14. We
also tried to vary the $J$ value according to the different scenarios
outlined above, but the fitting result depicted in Fig. \ref{fig:MvsH}
is very robust. However, the fitted $g$ value is decreasing from
2.14 to 1.2 for $J$ = 1/2 to $J$ = 3/2, respectively.

While the temperature-independent susceptibility contribution from
the CW fitting yields $\chi_{0}=5.83\times10^{-4}$\ cm$^{3}$/mol,
we obtain a larger value of $\chi_{0}=15.0\times10^{-4}$\ cm$^{3}$/mol
from the field dependence of the magnetization. This mismatch probably
arises from non-negligible antiferromagnetic spin correlations in
Ba$_{2}$YIrO$_{6}$ which have not been taken into account in the
Brillouin function. Further detailed investigations are planned in
order to shed light on the nature of the magnetic correlations in
Ba$_{2}$YIrO$_{6}$, which seem to increase at low temperatures.

\begin{table*}
\centering{}\protect\protect\caption{\label{tab:VanVleckSus} A comparison of the Van Vleck susceptibility
($\chi_{vv}$ ), effective magnetic moment ($\mu_{eff}$) and Curie-Weiss
temperature ($\theta_{\mathrm{CW}}$) of different $5d^{4}$ materials.}

\begin{tabular}{|c|c|c|c|c|c|}
\hline 
Material  & Electronic config.  & $\chi_{vv}$ (cm$^{3}$K/mol)  & $\mu_{eff}$ ($\mu_{\mathrm{B}}$/Ir) & $\theta_{\mathrm{CW}}$ (K) & Reference\tabularnewline
\hline 
\hline 
Ba$_{2}$YIrO$_{6}$  & $5d^{4}$  & $7.51\times10^{-4}$  & $0.44$ & $-8.9$ & This work\tabularnewline
\hline 
Sr$_{2}$YIrO$_{6}$  & $5d^{4}$  & $10.11\times10^{-4}$  & $0.91$ & $-229$  & \cite{Cao-PRL-2014-Sr2YIrO6}\tabularnewline
\hline 
NaIrO$_{3}$  & $5d^{4}$  & $19\times10^{-4}$  & $0.28$  & $-2.2$  & \cite{Bremholm-JSSC-184-2011}\tabularnewline
\hline 
\end{tabular}
\end{table*}

\subsection{Heat capacity}

The low-temperature specific heat ($C_{P}$) data for zero field is
shown on the left axis of Fig. \ref{fig:HC}. Our specific heat data
is qualitatively similar to that of Sr$_{2}$YIrO$_{6}$ \cite{Cao-PRL-2014-Sr2YIrO6}.
We do not find any signature of magnetic ordering down to $0.4$~K
in Ba$_{2}$YIrO$_{6}$, and no anomaly is resolved even in the $\frac{dC_{P}}{dT}$
vs $T$ plot, which is shown on the right axis of Fig. \ref{fig:HC},
in contrast to what has been reported for Sr$_{2}$YIrO$_{6}$.

\subsection{Ground state in DFT theory}

To understand the origin of the insulating ground state in Ba$_{2}$YIrO$_{6}$
we performed density functional theory (DFT) calculations of the electronic
structure. Our calculations were carried out within the local (spin)
density approximation {[}L(S)DA{]} using the Full Potential Local
Orbital band-structure package (FPLO) \cite{Koepernik,fplo}. A \textit{k}-mesh
of $12\times12\times12$ \textit{k}-points in the whole Brillouin
zone was employed. To take correlation effects in the Ir $5d$ shell
into account we adopted the L(S)DA+U scheme. Due to the rather sizable
spin-orbit interaction of the Ir atoms, the full relativistic four-component
Dirac scheme was used.

Similar to other iridates, the LDA results suggest a metallic state
for Ba$_{2}$YIrO$_{6}$ as indicated by a finite density of states
(DOS) at the Fermi level $E_{F}$ {[}Fig. \ref{fig:pDOS}(a){]}. The
Ir $5d$ and O $2p$ states have strong hybridization due to strong
metal-ligand covalency. The Ir $5d$ electrons give a contribution
to the total density of states in three energy windows: from $-6.5$\,eV
to $-4.5$\,eV, from $-1$\,eV to $0.5$\,eV, and from $2.5$\,eV
to $4.5$\,eV. From Fig. \ref{fig:pDOS} one can see that the $t_{2g}$
and $e_{g}$ states are well separated. The spin-orbit interaction
{[}Fig. \ref{fig:pDOS}(b){]} considerably changes the bands near
the Fermi level, but still the DOS at the Fermi level remains finite.
Analyzing the partial density of states, one can identify a splitting
of the $5d$ states into single-particle $5d_{3/2}$ and $5d_{5/2}$
contributions. To obtain an insulating ground state one needs to take
into account strong correlations in the mean-field approximation (LDA+U).
We introduce a Hubbard $U\approx1.4$~eV and Hunds $J=0.5$~eV for
the Ir $5d$ shell. These values of $U$ and $J$ give a magnetic
solution for the system with the magnetic moment of the order of $2\mu_{B}$
in the absence of the spin-orbit interaction. But with the spin-orbit
interaction the magnetic solution becomes unstable and instead we
find, at a configuration with an effective single particle, $J_{z}=S_{z}=0$,
as one can see from Figs. \ref{fig:pDOS}(c) and \ref{fig:bands}.
The presence of a moderate $U$ opens a gap $\Delta=0.2$ eV, pushing
up one of the bands of predominantly $5d_{5/2}$ character. This value
for the gap is close to one that we have found experimentally from
the resistivity measurements. From this we conclude that the insulating
state has a similar origin as in other iridates: the interplay of
spin-orbit interaction and correlations.

\section{Conclusions}

Single crystals of the double perovskite Ba$_{2}$YIrO$_{6}$ were
grown by the flux method. Our XRD measurements on single and crushed
crystals unambiguously reveal that this material crystallizes in a
cubic double perovskite structure. In contrast to the general expectation,
we found that Ba$_{2}$YIrO$_{6}$ is paramagnetic from our bulk susceptibility
measurements. The susceptibility data is fitted well with the CW formula
and results in an effective magnetic moment $\mu_{eff}=0.44$\ $\mu_{\mathrm{B}}$/Ir
and a Weiss temperature $\theta_{\mathrm{CW}}=-8.9$~K. However,
it is not clear at the moment if this is the manifestation of proposed
gapped excitonic magnetism \cite{Khaliullin-PRL-2013-d4magnetism,Meetei-arxiv-2013-d4magnetism}
in $d^{4}$ materials or caused by chemical disorder and / or off-stoichiometry
(the presence of Ir$^{4+}$ or Ir$^{6+}$ ions). Density-functional-based
electronic structure calculations show that in the LDA+U approach
a magnetic ground state is stable for physical values of the Hubbard
$U$ and Hund's rule exchange $J$, but only if relativistic effects
are treated on a scalar relativistic level (no spin-orbit coupling).
In fully relativistic calculations we find that the spin-orbit coupling
drives the system into a Mott insulator with a value of the gap close
to the experimental value. However, at the same time the system becomes
nonmagnetic. This calls for further investigations of the origin of
the unexpected magnetism in this material and suggests that the origin
of the observed magnetic moments is related to electronic many-body
effects, the theoretical description of which likely stretches beyond
the reach of effective mean-field approaches such as LDA+U.

\section{Acknowledgments}

We would like to thank F. Hammerath, D. Khomskii, and T. Saha Dasgupta
for fruitful discussion; S. Müller-Litvyani and J. Werner for technical
support; and L. Giebeler for support with the XRD data. S.W. acknowledges
funding by the Deutsche Forschungsgemeinschaft DFG under the Emmy-Noether
Programme (Project No. WU595/3-1) and by a Materials World Network
Grant (project WU595/5-1). O.K. acknowledges DFG support (Project
No. KN 393/20-1). This work has been supported by DFG in SFB~1143.


\begin{thebibliography}{10}
\bibitem{note1} For temperatures below 1.8~K only ZFC measurements
have been performed.
\end{thebibliography}
\end{document}